\RequirePackage{fix-cm}
\documentclass[smallextended]{svjour3}       
\smartqed  
\usepackage{graphicx}
\usepackage{color}
\begin{document}

\title{Three-particle Bell-like inequalities under Lorentz transformations}


\author{H. Moradpour  \and
        S. Maghool  \and S. A. Moosavi.
}


\institute{H. Moradpour \at
             Research Institute for Astronomy and Astrophysics of Maragha (RIAAM),
P.O. Box 55134-441, Maragha, Iran.\\
              Tel.: +984137412222\\
              Fax: +984137412224\\
              \email{h.moradpour@riaam.ac.ir}    \and
            S. Maghool, S. A. Moosavi \at
              Research Institute for Astronomy and Astrophysics of Maragha (RIAAM),
P.O. Box 55134-441, Maragha, Iran.}

\date{Received: date / Accepted: date}

\maketitle

\begin{abstract}
We study the effects of Lorentz transformations on three-particle
non-local system states (GHZ and W) of spin 1/2 particles, using
the Pauli spin operator and a three-particle generalization of
Bell's inequality, introduced by Svetlichny. In our setup, the
moving and laboratory frames used the (same) set of measurement
directions that maximally violate Svetlichny's inequality in the
laboratory frame. We also investigate the behavior of Mermin's and
Collins' inequalities. We find that, regardless of the particles'
type of entanglement, violation of Svetlichny's inequality in the
moving frame is decreased by increasing the boost velocity and the
energy of particles in the laboratory frame. In the relativistic
regime Svetlichny's inequality is a good criterion to investigate
the non-locality of the GHZ state. We also find that Mermin's and
Collins' inequalities lead to reasonable predictions, in agreement
with the behavior of the spin state, about non-locality of the W
state in the relativistic regime. Then, comparing our results with
those in which Czachor's relativistic spin is used instead of the
Pauli operator, we find that the results obtained by considering
the Pauli spin operator are in better agreement with the behavior
of spin state of the system in the relativistic information
theory. 
\end{abstract}
\maketitle
\bigskip

\section{Introduction \label{Introduction}}
Einstein, Podolsky and Rosen have found the non-local behavior of
multi-particle systems in the framework of Quantum Mechanics (EPR)
\cite{EPR}. Moreover, it is shown that the distinguishable
particles can behave non-locally \cite{BA}. In order to provide a
criterion for differing the local and non-local behaviors of
systems from each other, Bell introduced his inequality (the
Bell's inequality) which may be violated by the non-local
phenomena \cite{Bell}. The Bell's inequality is the backbone of
subsequent similar works \cite{CHSH,WIG,CH,rev}. Aspect and his
co-workers observed non-locality in their experiments
\cite{aspect1,aspect2,aspect3}. Also, it was proven that
one-particle systems can exhibit non-local behavior
\cite{vedral1,vedral2}.

Previously, it was thought that non-locality leads to
entanglement, i.e. the total state of the system of particles can
not be written as the product of states of its constitutes, and
therefore, the maximum violation of Bell's inequality is
obtainable for maximally entangled states including maximum
non-locality \cite{Bell,gisin91,gisinperes,pop92}. Hence, the
amount of entanglement, non-locality and the violation of Bell's
inequality were seen as the same. But, it is shown that there are
some mixtures of entangled states which do not violate any
two-partite Bell-type inequality \cite{pop92,werner89}. It is also
shown that two-qubit quantum states can be entangled without
violating any Bell-type inequality \cite{werner89,gisin96}.
Moreover, there are states which are not maximally entangled but
can maximally violate some types of Bell's inequality
\cite{eber,eber1,eber2,eber3,jun}. In fact, it is shown that the
amount of non-locality is not proportional to the amount of
entanglement, this means that it is possible to store a large
amount of non-locality into the states which are less entangled
\cite{eber4}. Indeed, unentangled systems may also exhibit the
quantum non-locality \cite{ben}. Therefore, the relation between
entanglement, non-locality and the violation of the Bell's
inequality is not as previously thought \cite{jun,gisin96,Bert}.

The entropy associated to non-locality has vast implications in
quantum information theory and its related topics
\cite{aud,Nilsen}. However, it is shown that the probability
distributions are frame dependant quantities \cite{wald} and
consequently, the entropy and information are frame dependent.
Peres and his colleagues confirmed the frame dependency for the
spin entropy of a single free spin-$\frac{1}{2}$ particle. They
have also shown that, due to the fact that the uncertainty in the
momentum state of the system will transport to the spin state, the
spin entropy is not invariant under the Lorentz transformations
(LT) \cite{peres1}. Therefore, we can conclude that in systems
with no uncertainty in their momentum states, the Wigner rotation
and thus the spin rotation in the moving frame are unique.
Therefore, the spin entropy does not change under LT. The similar
results hold for one-particle non-local systems \cite{vedral3}.
More debates on the results published by Peres and his colleagues
can be found in refs. \cite{peres2,peres3}. Additionally, it is
useful to mention here that relativistic motions may induce some
noises to the quantum cryptographic protocols
\cite{peres4,peres5}.

For high energy particles in the laboratory frame, QM is broken
down and we need a more comprehensive theory. In fact, under this
condition the particles have a non-negligible antiparticle
component \cite{wein}. Finally, in order to study the system, one
should either use the Relativistic Quantum Mechanics (RQM) or the
Quantum Field Theory (QFT) approaches, depending on the energy of
particles \cite{wald,wein}. Some attempts in which authors take
into account the QFT interpretation of phenomena, and study
non-locality can be found in
\cite{fun,smith,friis,moradi3,moradi4}.

Whenever the Quantum Mechanical interpretations of the phenomena
are satisfactory in the lab frame, then the system is
non-relativistic and the anti-particle component is negligible.
Here, we shall not consider any antiparticle production or
entanglement effects that could result from scattering. Therefore,
in our setup, the lab and moving frames are connected to each
other with a LT \cite{peres1,vedral3,hal,wigner}.

The effects of LT on the standard two-particle entanglement have
been extensively studied
\cite{gingrich,li,jordan,alsing,terashima,terashima1,ahn,lee,kim}.
In their approaches, there is a moving frame $(S^{\prime})$ which
is related to a lab frame $(S)$ by a LT. The results of the
generalizations of the frame dependency, shown by Peres et al.
\cite{peres1} and Vedral et al. \cite{vedral3} in one-particle
systems, to the two-particle pure entangled systems
\cite{gingrich,li,jordan} are analogous with the previous studies
on one-particle systems. Gravitational effects, the curvature of
spacetime \cite{fu,al,ma,le} and the acceleration
\cite{fun,smith,alsing,tera,shi,ball,ver}, do also change the
entanglement of the system.

Terashima et al. have considered the Pauli spin operator in order
to build the Bell operator, and studied the effects of LT on the
standard two-particle entangled systems (one of the Bell states),
in which the moving frame takes the same measurement directions
for the Bell operator as the lab frame. For their setup, Bell's
inequality is violated to its maximum value in the lab frame, and
the value of violation of Bell's inequality (non-locality), in the
moving frame, is a decreasing function of the boost speed and the
energy of the particles. Therefore in the moving frame, Bell's
inequality in the $\beta \rightarrow 1$ limit can be violated for
different energy levels of the particles in the lab frame
\cite{terashima,terashima1}.

Indeed, there are various operators suggested to describe the spin
of electron and therefore, the relativistic version of the
Stern-Gerlach experiment \cite{czachor,spin1,spin2,spin4}. Here,
since some authors used Czachor's relativistic spin operator in
order to study the relativistic version of EPR
\cite{ahn,lee,kim,czachor,spin4,moradi1,mm,moradi2}, and because
we will point to the results of considering Czachor's relativistic
spin operator, it is useful to introduce this spin operator. Based
on Czachor's proposal \cite{czachor}, the relativistic spin
operator of states with zero momentum uncertainty along the unit
vector $\overrightarrow{A}$ is
\begin{eqnarray}
\widehat{A}=\frac{(\sqrt{1-\beta^2}\overrightarrow{A}_{\bot}
+\overrightarrow{A}_{\|}).\overrightarrow{\sigma}}{\sqrt{1+\beta^2[(\widehat{e}
.\overrightarrow{A})^2-1]}}.
\end{eqnarray}
In this equation, $\sigma$ and $\widehat{e}$ are the Pauli spin
operator and the unit vector along the particle velocity direction
respectively, whiles, $\beta$ is the particle velocity. In
addition, the subscripts $\bot$ and $\|$ denote the perpendicular
and the parallel components of the vector $\overrightarrow{A}$ to
the direction of the particle velocity. This operator commutes
with the Hamiltonian and covers the Pauli spin operator whenever
$\overrightarrow{A}_{\|}=0$ meaning that
$\widehat{A}=\overrightarrow{A}.\widehat{\sigma}$. We should note
that the uncertainty principle inhibits such possibility in a
realistic experiment \cite{czachor}. More properties of the
Czachor's relativistic spin operator can be found in
\cite{czachor,spin1,spin2,spin4}. The generalization of this
operator to the cases in which we deal with wave-packets instead
of a plane wave can be found in \cite{peres5}.

Now consider a setup in which the lab frame uses Czachor's
relativistic spin operator to build the Bell operator, and the
special direction for measuring the spin operator leading to
$\widehat{A}=\overrightarrow{A}.\widehat{\sigma}$, which is
similar to the result of considering the Pauli operator
\cite{terashima,czachor}. Therefore, Bell's inequality will be
violated to its maximum value in the lab frame by choosing proper
directions for $\overrightarrow{A}$ \cite{ahn,lee,kim,czachor}.
Thereinafter, consider another frame moving with respect to the
lab frame along an axis which is not parallel to the particles
velocity in the lab frame while the spin measurement in the moving
frame is also directed along $\overrightarrow{A}$. It means that
$\overrightarrow{A}_{\|}$ is generally non-zero in the moving
frame \cite{ahn,lee,kim,czachor}. Therefore, since particles have
an extra motion (along the boost direction) in the moving frame
compared with the lab frame, $\overrightarrow{A}_{\bot}$ in the
moving frame differs from that of the lab frame. Briefly, since
the moving frame uses the same direction as the lab frame for
measuring the spin operator ($\overrightarrow{A}$),
$\overrightarrow{A}_{\|}$ and $\overrightarrow{A}_{\bot}$ differ
from those of the lab frame and
$\widehat{A}\neq\overrightarrow{A}.\widehat{\sigma}$ in the moving
frame. Finally, independent of the energy of the particles, the
value of violation of Bell's inequality decreases as a function of
the boost velocity in the moving frame and eventually, Bell's
inequality will be served in the $\beta \rightarrow 1$ limit
\cite{ahn,lee,kim}. We should mention here that the maximum
violation of Bell's inequality is also obtainable in the moving
frame by choosing proper directions for $\overrightarrow{A}$ in
the moving frame \cite{kim}.

Therefore, it is apparent that the results of studying Bell's
inequality by using Czachor's relativistic spin operator differ
from those in which the Pauli operator is used in order to build
the Bell operator. Here, we must note that if there is no
uncertainty in the momentum state of each particle in two-particle
systems then the maximum violation of Bell's inequality in the
moving frame will be accessible by simultaneous application of the
LT on the quantum state and the Bell operator
\cite{terashima,terashima1,kim}.

There are two well known three-particle pure entangled states
including GHZ and W states. The GHZ and W states are written as
\begin{eqnarray}\label{cf}
|GHZ\rangle = \frac{1}{\sqrt{2}}(|+++\rangle +|---\rangle)\
\end{eqnarray}
and
\begin{eqnarray}
|W\rangle=\frac{1}{\sqrt{3}}(|++-\rangle+|+-+\rangle+|-++\rangle),
\end{eqnarray}
which include the genuine three-partite entanglement
\cite{genu,genu1,dur,Mitchel,cerc,svet,svet1}. On one hand, both
of them are genuine three-partite entangled states while on the
other hand, the GHZ state is separable by measuring the spin of
one particle. Whereas a spin measurement on the W state leads to a
separable state with probability equal to $\frac{1}{3}$. Indeed, a
spin measurement on the W state turns this state into the
two-partite entangled state
\begin{eqnarray}
|B\rangle=\frac{1}{\sqrt{2}}(|+-\rangle+|-+\rangle),
\end{eqnarray}
with probability equal to $\frac{2}{3}$. This remaining part
includes the spin state of the other particles. This difference
leads to different values for the $3$-tangle measure ($\tau$) of
entanglement. Indeed, whiles $\tau$ vanishes for the W state, its
value is non-zero for the GHZ state \cite{dur,cofman}. Therefore,
one can use the $3$-tangle measure to distinguish the GHZ and W
states. Moreover, one can always convert nontrivial three-particle
entangled states into one of these states by stochastic local
operations and classical communication (SLOCC) whiles these two
classes of states cannot be converted to each other by a SLOCC
\cite{dur,cerc}. These arguments show that although both of the W
and GHZ states include genuine three-partite non-locality
\cite{genu,genu1,dur,Mitchel,cerc,svet,svet1}, but their
entanglement type differ from each other \cite{genu1}. More
relations between the W and GHZ states and three-partite entangled
states can be found in \cite{genu,genu1,Mitchel,svet}.

Three-particle version of Svetlichny's inequality $(|S_v|)$, as
the generalization of Bell's inequality to the multi-particle
systems, is written as \cite{svet,svet1}
\begin{eqnarray}\label{svet}
&|S_v|=|E(ABC)+E(ABC')+E(AB'C)+E(A'BC)-E(A'B'C')\nonumber
\\ &-E(A'B'C)-E(A'BC')-E(AB'C')|\leq 4,
\end{eqnarray}
where we have
\begin{eqnarray}\label{cf}
&E_{GHZ}(ABC)=\langle GHZ|\sigma(\widehat{n_1})\otimes
\sigma(\widehat{n_2}) \otimes
\sigma(\widehat{n_3})|GHZ\rangle\nonumber
\\ &=\cos(\phi_1+\phi_2+\phi_3),
\end{eqnarray}
and
\begin{eqnarray}\label{cf2}
&E_{W}(ABC)=\langle W|\sigma(\widehat{N_1})\otimes
\sigma(\widehat{N_2})\otimes \sigma(\widehat{N_3}) |W\rangle
\nonumber
\\ &=-\frac{2}{3}\cos(\theta_1+\theta_2+\theta_3)-\frac{1}{3}\cos\theta_1
\cos\theta_2 \cos\theta_3.
\end{eqnarray}
In the above equations, $A$ and $A'$ are possible measurements on
the first particle and the same relations hold for the second and
third particles with possible measurements $B$, $B'$ and $C$,
$C'$, respectively. $\widehat{n_i}$ and $\widehat{N_i}$ are the
unit vectors in the $x-y$ and $x-z$ planes, respectively.
$\widehat{n_i}$ is characterized by the azimuthal angle $\phi_i$,
and the maximum possible violation ($4\sqrt{2}$) of $|S_v|$ for
the GHZ state will be obtainable if $\Sigma \phi_i =
(n+\frac{3}{4})\pi$ and $\Sigma \phi'_i = (n+\frac{9}{4})\pi$,
where $(n=0,\pm1,\pm2,...)$ \cite{cerc}. For the W state, $|S_v|$
will be at most violated to the value $4.354$ if $\theta_i$ (the
polar angle), which specifies the unit vector $\widehat{N_i}$,
satisfies the condition $\theta_i=\pi-\theta'_i=35.264^{\circ}$
\cite{cerc}. Loosely speaking, since the maximum violation value
of the $|S_v|$ inequality for the GHZ state ($4\sqrt{2}$) is
greater than that of the W state ($4.354$), the same as the
$3$-tangle measure, one can also use this inequality to
distinguish the GHZ and W states \cite{cerc,genu1}.

While, the GHZ and W states can violate the $|S_v|$ inequality
\cite{smith,genu,genu1,Mitchel,cerc,svet}, three-particle systems
with bi-partite non-locality cannot violate the $|S_v|$ inequality
\cite{genu,genu1,svet}. Therefore, Svetlichny's inequality can
also be used to distinguish the bi-partite non-locality and the
non-locality stored in the GHZ and W states
\cite{genu,genu1,Mitchel,cerc}.

There are also two other Bell-like inequalities for $n$-particle
systems derived by Mermin $(M)$ and Collins et al. $(M')$ which in
the three-particle case are written as \cite{mermin,cal}:
\begin{eqnarray}\label{mer}
|M|&=&|E(ABC')+E(AB'C)+E(A'BC)-E(A'B'C')|\leq 2, \nonumber \\
|M'|&=&|E(ABC)-E(A'B'C)-E(A'BC')-E(AB'C')|\leq 2.
\end{eqnarray}
One gets the values $4$ and $3.046$ as the maximum amount of
violation of these inequalities for the GHZ and W states
respectively \cite{cerc}. Therefore, these inequalities can also
be used to distinguish the GHZ and W states. Gisin et al. have
shown that only the GHZ state can violate $|M|$ and $|M'|$ to the
values greater than $2\sqrt{2}$ \cite{gisin}. Here, it is useful
to mention that there is no set of measurements violating $|M|$
and $|M'|$ to the values greater than $2\sqrt{2}$ simultaneously
\cite{cerc}. Finally, we should note that since the maximum
violation amounts of the three-particle Bell-like inequalities for
the GHZ state are greater than that of the W state, the three
particles are more entangled in the GHZ state than the W state
\cite{cerc}. The latter may be supported by this fact that the
value of $3$-tangle measure for these states differs from each
other \cite{dur,cofman}.

When $|S_v|$ is violated to its maximum value, for the $|M|$ and
$|M'|$ inequalities, one will find
\begin{eqnarray}
|M|=|M'|=\frac{|S_v|}{2}.
\end{eqnarray}
Therefore, if we use the set of measurements that violate $|S_v|$
to its maximum value, then $|M|$ and $|M'|$ will also be violated
\cite{cerc}. But, since there is no set of measurements that
simultaneously violate the $|M|$ and $|M'|$ inequalities to the
amounts greater than $\frac{|S_v|}{2}$, the violations of $|M|$
and $|M'|$ to their maximum violation amounts ($4$ and $3.046$ for
the GHZ and W states, respectively) cannot happen simultaneously
\cite{cerc}. In addition, If we consider the GHZ state and the
special set of measurement angles violating $|S_v|$ to the value
$4\sqrt{2}$, we will find $|M|=|M'|=2\sqrt{2}$ which according to
Gisin et al. \cite{gisin}, we can conclude that there is no
genuine three-partite entanglement in the GHZ state. This result
is fully inconsistent with some attempts which claim that the GHZ
state includes genuine three-partite entanglement
\cite{genu,genu1,dur,Mitchel,cerc,svet,svet1}. This example shows
that the $|M|$ and $|M'|$ inequalities are weaker criteria than
the $|S_v|$ inequality in order to study the GHZ state
\cite{cerc}. We also see that the dependency of the three-particle
generalizations of Bell's inequality to the set of measurements is
extremely more than that of Bell's inequality itself.

It seems that, in non-relativistic regimes, one can use $|S_v|$ as
the proper three-particle Bell-like inequality in order to study
the GHZ and W states \cite{Mitchel,cerc,svet,smith}. This is of
course due to the following points: ($i$) $|M|$ and $|M'|$ are
violated by a hybrid local-nonlocal hidden variables model, and
the same situation holds for those inequalities which include four
correlation functions $(E(ABC))$ \cite{cerc,cal}. ($ii$) It is
shown that the upper bound of Mermin's inequality is not correct
and should be revised \cite{roy}. ($iii$) It has been shown that
if $|M|$ is violated to its maximum value, then $|M'|$ will be
satisfied and vice versa \cite{cerc}. In addition, for the GHZ
state, there is no set of the measurement angles that violate
$|M|$ and $|M'|$ to the values bigger than $2\sqrt{2}$
simultaneously \cite{cerc}. Therefore, it seems that we will
confront a contradiction if we compare the results of $|M|$ and
$|M'|$ either with each other or that of $|S_v|$. ($iv$) The $|M|$
and $|M'|$ inequalities are weaker than what is needed to study
the multi-partite non-locality \cite{AA}. ($v$) The violation
amounts of $|S_v|$ for the GHZ and W states are different and
therefore $|S_v|$ can distinguish these states
\cite{Mitchel,cerc}.

We must note that the correlation functions in $|S_v|$ are
stronger than those needed for detecting the non-locality stored
in the GHZ and W states \cite{AA}, but it seems that $|S_v|$ can
detect these non-localities, and one can use $|S_v|$ in the study
of the non-relativistic systems \cite{Mitchel,cerc,rev}. In fact,
the $|S_v|$ inequality is offended from the violation of the
no-signalling constraint \cite{gw,bp,ac}, but this shortcoming of
$|S_v|$ can be eliminated by either considering bi-partite
correlations which satisfy the no-signalling constraint or using
the time-ordered bi-partite correlation functions \cite{bp,rev}.
In fact, such corrections eliminate the possibility of forming the
grandfather-type paradoxes. For a comprehensive review on this
subject, see ref.~\cite{rev} and references therein.

Authors in ref.~\cite{you} have studied the effects of LT on the
GHZ state. They have used the $|M|$ inequality, considering the
Pauli spin operator, and the special set of measurement angles
that violate $|M|$ to its maximum value ($4$) in the lab frame.
They have also used the same measurement angles for the moving
frame in order to evaluate $|M|$. In addition, they have shown
that the violation amount of the $|M|$ inequality in the moving
frame is a decreasing function of the boost velocity and the
energy of the particles. Finally, they found that, in the $\beta
\rightarrow 1$ limit, the violation amount of $|M|$ depends on the
energy of particles in the lab frame. These results are in line
with previous studies on the two-particle non-local systems
\cite{terashima,terashima1}. Therefore, we may conclude that the
behavior of non-locality under LT is independent of the number of
particles.

In a similar setup, Moradi \cite{moradi1} has considered Czachor's
relativistic spin operator instead of the Pauli operator in order
to evaluate the $|M|$ inequality. He found that the violation
amount of the $|M|$ inequality in the moving frame, in the
$\beta\rightarrow1$ limit, depends on the energy of the particles
in the lab frame, which is against the two-particle entangled
system studies \cite{ahn,lee,kim}. Therefore, it seems that the
behavior of non-locality under LT depends on the number of
particles, and it is independent of the nature of non-locality.
This result is fully inconsistent with the previous result by You
et al. \cite{you}. Due to the fact that the authors in
references~\cite{you} and~\cite{moradi1} have used the $|M|$
inequality in order to study the GHZ state, their results are
doubted. Bearing the differences between the GHZ and W states in
mind, since authors in \cite{you,moradi1} did not consider the W
state, one can not generalize their results to the W state.

It has been shown that, in the moving frame, by considering the
GHZ and W states and the Czachor's relativistic spin operator to
evaluate three-particle Bell-like inequalities ($|M|$, $|M'|$ and
$|S_v|$), and also, using the same special set of measurements
that violate $|S_v|$ to its maximum value in the lab frame, these
inequalities will be satisfied in the $\beta\rightarrow 1$ limit,
independent of the energy of particles in the lab frame \cite{mm}
which is in line with two-particle studies \cite{ahn,lee,kim}.
Therefore, in the $\beta\rightarrow 1$ limit, Bell's inequality
and multi-particle Bell-like inequalities will be satisfied under
LT independent of the number of the entangled particles, their
energies in the lab frame and type of their entanglement, if and
only if the moving observer uses Czachor's relativistic spin
operator as well as the special set of measurements that violate
either $|S_v|$ or Bell's inequality, depending on the considered
system, to its maximum value in the lab frame. Finally, it seems
that the $|S_v|$ inequality is a good witness for detecting
non-locality in relativistic multi-particle systems \cite{mm}.

It is useful to note that non-local systems, it was shown that the
maximum violation of Bell's inequality and its generalization to
the three-particle systems in the non-accelerated moving frame is
accessible if one applies LT on the spin state and the spin
operator simultaneously
\cite{terashima,terashima1,kim,moradi1,moradi2,you}.

Moreover, the acceleration effects on two and three-particle
entangled states have also been studied
\cite{smith,friis,hwang,wang}. It is shown that the $|S_v|$
inequality can be violated for any finite value of acceleration
whiles Bell's inequality cannot be violated for sufficiently large
but finite acceleration. Therefore, it seems that the effects of
acceleration on two-particle entanglement differs from those of
three-particle entangled states such as the GHZ state
\cite{smith}. It is useful to note that the $|S_v|$ inequality is
also satisfied for infinite value of acceleration
\cite{smith,friis}.

Here, we study the behavior of three-particle non-local systems,
which either include the GHZ state or the W state, under LT. In
order to investigate the behavior of multi-particle Bell-like
inequalities under LT we use the Pauli spin operator and the same
set of measurements for the moving and lab frames that violate the
$|S_v|$ inequality to its maximum value in the lab frame. We also,
compared the results obtained from using $|M|$ and $|M'|$ with
those obtained from using the $|S_v|$ inequality. We show that, in
the relativistic regime, the inequalities using the Pauli spin
operator instead of Czachor's operator to study the entanglement
are in more agreement with the behavior of the quantum mechanical
spin states transformed by the LT. Throughout this paper we set
the light velocity equal to one ($c=1$) for simplicity.

The paper is organized as follows. In section ($\textmd{II}$), we
introduce the Wigner rotation, consider the Pauli spin operator,
and show that the $|S_v|$ inequality is a good witness for
studying the GHZ state. Thereinafter, we consider the W state, and
point to the weakness of $|S_v|$ compared with the $|M|$ and
$|M'|$ inequalities in the relativistic regime. Finally, we find
that the $|M|$ and $|M'|$ inequalities can lead to reasonable
predictions about the behavior of the GHZ and W states under LT.
Throughout the article, the results of considering Czachor's
operator are also addressed. Section ($\textmd{III}$) is devoted
to a summary and concluding remarks.
\section{The three particle non-local system under LT}
Here, we consider a situation in which Quantum Mechanics is enough
for describing the system in the lab and thus moving frames. This
situation is considered by many authors for both of the low and
high energy particles, and some of these efforts can be found in
refs.
\cite{peres1,vedral3,peres2,peres3,peres4,peres5,gingrich,li,jordan,alsing,terashima,terashima1,ahn,lee,kim,moradi1,mm,moradi2}.
In the lab frame ($S$) for a system, including three
spin-$\frac{1}{2}$ particles with the spin state
$\vert\psi\rangle$ and the momentum state $\vert
\overrightarrow{p_1}\overrightarrow{p_2}\overrightarrow{p_3}\rangle$,
the state of the system is written as
\begin{eqnarray}
\vert \xi\rangle=\vert
\overrightarrow{p_1}\overrightarrow{p_2}\overrightarrow{p_3}\rangle\vert\psi\rangle,
\end{eqnarray}
where $\overrightarrow{p_i}=p_0 \hat{z}$, $\forall~i$. Now,
consider a moving frame ($S'$) which moves along the $x$ axis
$(\overrightarrow{\beta}=\beta\widehat{x})$. In the $S'$ frame,
the state of the system is
\begin{eqnarray}
\vert \xi \rangle^{\Lambda}=\vert
\overrightarrow{p_1}\overrightarrow{p_2}\overrightarrow{p_3}
\rangle^{\Lambda} \prod_{i=1}^{3} D(W(\Lambda,p_i))
\vert\psi\rangle.
\end{eqnarray}
$\vert \overrightarrow{p_1}\overrightarrow{p_2}\overrightarrow{p_3}
\rangle^{\Lambda}$ denotes the momentum state of the system in the
moving frame, and $D(W(\Lambda,p_i))$ is the spin-$\frac{1}{2}$
Wigner representation of the Lorentz group for the i$^{\textmd{th}}$
particle \cite{hal,wigner}:
\begin{eqnarray}\label{wr}
D(W(\Lambda,p_i))=\cos\frac{\Omega_{p_i}}{2}-i\sigma_y
\sin\frac{\Omega_{p_i}}{2}.
\end{eqnarray}
In this equation, $\sigma_y$ is the Pauli matrix where
$\Omega_{p_i}$ is called the Wigner angle and will be evaluated as
\begin{eqnarray}\label{angle}
\tan\Omega_p=\frac{\sinh\alpha\sinh\delta}{\cosh\alpha +
\cosh\delta}.
\end{eqnarray}
Here, $\cosh\delta=\frac{p_0}{m}$ and
$\cosh\alpha=\sqrt{1-\beta^2}$.
\subsection{the GHZ state}
In the laboratory frame, the entangled particles are in the GHZ
state. Thus
\begin{eqnarray}
|\psi \rangle=|GHZ\rangle,
\end{eqnarray}
and the maximum violation of the $|S_v|$ inequality ($4\sqrt{2}$)
in the $S$ frame is obtainable by using the special set of angles
including $\phi_i=\frac{\pi}{4}$ and $\phi'_i=\frac{3\pi}{4}$.
This set of measurements yield $|M|=|M'|=2\sqrt{2}$, as we have
pointed in the introduction, which indicates that the system does
not contain the genuine three-partite entanglement \cite{gisin}.
Also, we know that this is not true for the GHZ state because the
GHZ state includes the genuine three-partite entanglement
\cite{genu,genu1,dur,Mitchel,cerc,svet,svet1}. Therefore, we see
again that this example can clarify our assertion about the more
sensitivity of the three-particle inequalities to measurements
with respect to Bell's inequality \cite{cerc}. It is useful to
remind that since there is no set of measurements that
simultaneously violate $|M|$ and $|M'|$ to the values bigger than
$2\sqrt{2}$, the maximum violation amounts of $|M|$ and $|M'|$
cannot be obtained simultaneously \cite{cerc}.

The spin state in the moving frame,
($|GHZ\rangle^{\Lambda}~\equiv\prod_{i=1}^3 D(W(\Lambda,p_i))\vert
GHZ\rangle$), is written as
\begin{eqnarray}\label{ghz}
|GHZ\rangle^{\Lambda}&=&(\cos(\frac{\Omega_p}{2}))^3\vert
GHZ\rangle+(\sin(\frac{\Omega_p}{2}))^3\vert \overline{GHZ}\rangle
\\
&+&\sqrt{\frac{3}{2}}\sin(\frac{\Omega_p}{2})\cos(\frac{\Omega_p}{2})
[(\sin(\frac{\Omega_p}{2})+\cos(\frac{\Omega_p}{2}))\vert
W\rangle\nonumber \\
&+&(\sin(\frac{\Omega_p}{2})-\cos(\frac{\Omega_p}{2}))\vert
\overline{W}\rangle],\nonumber
\end{eqnarray}
where $\vert\overline{GHZ}\rangle=\frac{1}{\sqrt{2}}(\vert
---\rangle - \vert +++\rangle)$
and $\vert\overline{W}\rangle=\frac{1}{\sqrt{3}}(\vert --+\rangle
+\vert -+-\rangle + \vert +--\rangle)$. Simple calculations yield
\begin{eqnarray}\label{wbar}
&E_{\overline{W}}(ABC)=\langle
\overline{W}|\sigma(\widehat{N_1})\otimes
\sigma(\widehat{N_2})\otimes \sigma(\widehat{N_3}) |
\overline{W}\rangle \nonumber
\\ &=-\frac{2}{3}\cos(\theta_1+\theta_2+\theta_3)-\frac{1}{3}\cos\theta_1
\cos\theta_2 \cos\theta_3
\end{eqnarray}
and
\begin{eqnarray}\label{gbar}
&E_{\overline{GHZ}}(ABC)=\langle
\overline{GHZ}|\sigma(\widehat{n_1})\otimes
\sigma(\widehat{n_2})\otimes \sigma(\widehat{n_3}) |
\overline{GHZ}\rangle\nonumber
\\ &=-\cos(\phi_1+\phi_2+\phi_3).
\end{eqnarray}
In these equations, $\widehat{n_i}$ and $\widehat{N_i}$ have the
same definitions as those of the GHZ and W states respectively. In
addition, These equations tell that every set of measurements used
to detect the non-locality in either the W state or the GHZ state
can also be used for $\vert \overline{W}\rangle$ and $\vert
\overline{GHZ}\rangle$ respectively. Using Eq.~(\ref{cf}) for
correlation functions $E(ABC)$, we get
\begin{eqnarray}\label{ghzcr1}
&E&(ABC)=~^{\Lambda}\langle GHZ|\sigma(\widehat{n_1})\otimes
\sigma(\widehat{n_2}) \otimes
\sigma(\widehat{n_3})|GHZ\rangle^{\Lambda}=[(\cos(\frac{\Omega_p}{2}))^6
\nonumber
\\ &-&(\sin(\frac{\Omega_p}{2}))^6]\cos(\varphi_1 + \varphi_2 +
\varphi_3)-\frac{3}{4}(\sin\Omega_p)^2\cos\Omega_p[\cos(\varphi_1
+ \varphi_2 - \varphi_3)\nonumber
\\ &+&\cos(\varphi_1 - \varphi_2 +
\varphi_3)+\cos(-\varphi_1 + \varphi_2 + \varphi_3)],
\end{eqnarray}
which is compatible with Eq.~(\ref{cf}) in the $\Omega_p=0$ limit.
Inserting $\phi_i=\frac{\pi}{4}$ and $\phi'_i=\frac{3\pi}{4}$ into
the above equation, we find
\begin{eqnarray}\label{ghzcr}
&E&(ABC)=(-\frac{\sqrt{2}}{2})[(\cos(\frac{\Omega_p}{2}))^6-
(\sin(\frac{\Omega_p}{2}))^6]-\frac{9\sqrt{2}}{8}(\sin\Omega_p)
^2\cos\Omega_p\nonumber
\\ &=&-E(A'B'C'),\nonumber\\
&E&(A'BC')=E(AB'C')=E(A'B'C)=(\frac{\sqrt{2}}{2})[(\cos
(\frac{\Omega_p}{2}))^6-(\sin(\frac{\Omega_p}{2}))^6]\nonumber
\\ &-&\frac{3\sqrt{2}}{8}(\sin\Omega_p)^2\cos\Omega_p, \\
&E&(A'BC)=E(AB'C)=E(ABC')=-E(A'B'C).\nonumber
\end{eqnarray}
For the $|M|$, $|M'|$ and $|S_v|$ inequalities, we get
\begin{eqnarray}\label{vg}
|M|=|M'|=\frac{|S_v|}{2}=
\vert-2\sqrt{2}((\cos(\frac{\Omega_p}{2}))^6-(\sin(\frac{\Omega_p}{2}))^6)\vert.
\end{eqnarray}
It is straightforward that, in the $\Omega_p\rightarrow0$ limit,
the results of the $S$ frame are accessible, as a desired result.
In addition and for the ultra relativistic regimes
($\beta\rightarrow1$), calculations lead to
\begin{eqnarray}\label{lvg}
|M|=|M'|=\frac{|S_v|}{2}\sim \frac{1+3\Gamma^2}{\sqrt{2}\Gamma^3},
\end{eqnarray}
where $\Gamma=\frac{1}{\sqrt{1-v_0^2}}$ and $v_0$ is the energy
factor and the velocity of particles in the $S$ frame
respectively. We have also used the approximation,
$\sin\Omega_p\sim\sqrt{1-\frac{1}{\Gamma^2}}$ to derive
Eq.~(\ref{lvg}).

In this limit
$\sin\frac{\Omega}{2}\sim\sqrt{\frac{\Gamma-1}{2\Gamma}}$ and from
Eq.~(\ref{ghz}), we get
\begin{eqnarray}\label{ghz1}
|GHZ\rangle^{\Lambda}\sim|GHZ\rangle,
\end{eqnarray}
for low energy particles ($\Gamma\rightarrow1$), and
\begin{eqnarray}\label{ghz2}
|GHZ\rangle^{\Lambda}\sim\frac{1}{2}(|---\rangle+\sqrt{\frac{3}{2}}|W\rangle),
\end{eqnarray}
for high energy particles ($\Gamma\rightarrow\infty$). In
addition, for the high energy particles in the $\beta\rightarrow1$
limit from Eq.~(\ref{ghzcr}) we get
\begin{eqnarray}
E(ABC)\sim 0.
\end{eqnarray}

Eq.~(\ref{lvg}) shows that in the $\beta\rightarrow1$ limit:
($\textmd{i}$) the high energy particles ($\Gamma\rightarrow
\infty$) satisfy all of the inequalities which is compatible with
Eq.~(\ref{ghz2}). ($\textmd{ii}$) in accordance with
Eq.~(\ref{ghz1}), the inequalities will be violated to their
violation value in the $S$ frame for the low energy particles
($\Gamma\rightarrow 1$). Our result is in line with the previous
studies on the standard two-particle entanglement (Bell states)
\cite{terashima,terashima1}. Therefore, we see that the violation
of the inequalities in the $\beta\rightarrow1$ limit depends on
the energy of the particles in the $S$ frame. This result is
compatible with the previous studies by You et al. \cite{you} and
in line with the Moradi's calculations, in which Czachor's
relativistic spin operator are used instead of the Pauli operator,
\cite{moradi1,moradi2}. It is useful to note that since we assumed
that we work in the Quantum Mechanic framework, the high energy
limit ($\Gamma\rightarrow \infty$) is problematic. Indeed, Quantum
Mechanics has no desired efficiency in this limit, and we need to
consider more comprehensive theories such as RQM and QFT
\cite{wein}. Finally, we think that the survey of this limit may
lead to useful outcomes about the quality of consistency between
Quantum Mechanics and LT (Special Relativity) which may lead to
the some desired predictions about the effects of relative motion
on the Quantum Mechanical phenomena such as the relativistic
version of the Stern-Gerlach experiment
\cite{hal,terashima,czachor}.

In addition, by using Czachor's relativistic spin operator instead
of the Pauli operator one gets
\begin{eqnarray}\label{seven}
|M|=|M^{\prime}|=\frac{|S_v|}{2}=\frac{2|\cos\Omega_p
|}{\sqrt{2-\beta^2}^3}(\cos^2\Omega_p +3(1-\beta^2)),
\end{eqnarray}
which leads to
\begin{eqnarray}\label{seven1}
|M|=|M^{\prime}|=\frac{|S_v|}{2}\sim\frac{2}{\Gamma^3},
\end{eqnarray}
in the ultra relativistic limit ($\beta\rightarrow1$) \cite{mm}.
Eqs.~(\ref{seven}) and~(\ref{seven1}) indicate that, independent
of the energy of the particles in the lab frame, the inequalities
are satisfied in the $\beta\rightarrow1$ limit which is
inconsistent with the results of Eqs.~(\ref{vg}) and~(\ref{lvg}).
Although, Eqs.~(\ref{seven}) and~(\ref{seven1}) are in line with
previous studies on the Bell states \cite{ahn,lee,kim} but, they
have full contradiction with the asymptotic behavior of the spin
state of system (Eq.~(\ref{ghz1}) and~(\ref{ghz2})).

As we have mentioned in the introduction, since authors in
\cite{moradi1,moradi2,you} have used the $|M|$ inequality in order
to study the behavior of the GHZ state under LT, their results are
doubted. Here, we used the $|S_v|$ inequality and get the similar
results as obtained in the references
\cite{terashima,terashima1,moradi1,moradi2,you}. Therefore, based
on Eqs.~(\ref{vg}) and~(\ref{seven}) and the asymptotic behavior
of the spin state in the moving frame (Eqs.~(\ref{ghz1})
and~(\ref{ghz2})), we think that the results of studying the
effects of LT on the non-locality of GHZ state will be compatible
with the LT of this Quantum Mechanical state if the Pauli spin
operator, the $|S_v|$ inequality and the special set of
measurements, violating $|S_v|$ to its maximum violation amount
($4\sqrt{2}$), are considered. In addition, bearing the
difficulties of the $|S_v|$, $|M|$ and $|M'|$ inequalities in
mind, our investigation shows that the inequalities with
correlation functions stronger than those of $|M|$ and $|M'|$, and
weaker than those of the $|S_v|$ inequality may be used to study
the effects of LT on the GHZ state which is in line with
non-relativistic study \cite{AA}.

We should note that since there is no uncertainty in the momentums
of the particles \cite{peres1,vedral3,gingrich,li,jordan}, if the
moving observer applies LT on both of the spin operators and the
quantum state of the system simultaneously, the maximum violation of
the inequalities will be obtainable in the $S'$ frame
\cite{terashima,terashima1,kim,moradi1,mm,moradi2,you}.
\subsection{The W state}
In the lab frame, consider a situation in which the W state is the
spin state of particles which move along the $z$ direction with
the same momentum. In the moving frame, which moves along the $x$
direction with $\overrightarrow{\beta}=\beta\widehat{x}$, the
Wigner rotation satisfies Eq.~(\ref{wr}) and by following the
procedure of the previous section for the W state we get
\begin{eqnarray}
\vert W \rangle^\Lambda
&=&\sqrt{3}\sin(\frac{\Omega_p}{2})\cos(\frac{\Omega_p}{2})[-\cos(\frac{\Omega_p}{2})\vert+++\rangle
+\sin(\frac{\Omega_p}{2}) \vert---\rangle]\nonumber
\\&+&[(\cos(\frac{\Omega_p}{2}))^3-2\cos(\frac{\Omega_p}{2})
(\sin(\frac{\Omega_p}{2}))^2]\vert W\rangle\nonumber
\\&+&[2\sin(\frac{\Omega_p}{2})
(\cos(\frac{\Omega_p}{2}))^2-(\sin(\frac{\Omega_p}{2}))^3] \vert
\overline{W}\rangle,
\end{eqnarray}
where $\vert W \rangle^\Lambda$ is the spin state in the moving
frame. In addition, calculations for the correlation function lead
to
\begin{eqnarray}
E_W(\theta_1\theta_2\theta_3)&=& A \cos\theta_1 \cos\theta_2
\cos\theta_3+B \sin\theta_1 \sin\theta_2 \sin\theta_3\nonumber
\\&+&
C\cos(\theta_1 +\theta_2 +\theta_3)+D(\cos\theta_1
\sin\theta_2 \cos\theta_3\nonumber \\
&+&\sin\theta_1 \cos\theta_2 \cos\theta_3 + \cos\theta_1
\cos\theta_2 \sin\theta_3),
\end{eqnarray}
where we have used these abbreviations
\begin{eqnarray}
A&=&-\frac{1}{3}(\cos\Omega_p)^3+\frac{7}{6}(\sin\Omega_p)^2\cos\Omega_p,\nonumber
\\
B&=&\sin\Omega_p[ 2(\cos\Omega_p)^2 - (\sin\Omega_p)^2],\nonumber \\
C&=&\frac{7}{3} (\sin\Omega_p)^2 \cos\Omega_p -
\frac{2}{3}(\cos\Omega_p)^3,\nonumber \\
D&=&\sin\Omega_p[\frac{-7}{3}(\cos\Omega_p)^2+\frac{2}{3}(\sin\Omega_p)^3].
\end{eqnarray}
In the non-relativistic regime $(\Omega_p=0)$ we get
\begin{eqnarray}
E_W(\theta_1\theta_2\theta_3)&=&
-\cos\theta_1\cos\theta_3\cos\theta_3
+\frac{2}{3}(\cos\theta_1\sin\theta_2\sin\theta_3\nonumber
\\&+&\cos\theta_2\sin\theta_1\sin\theta_3
+\cos\theta_3\sin\theta_2\sin\theta_1),
\end{eqnarray}
which is the same as Eq.~(\ref{cf2}). Now, by inserting
$\theta_i=\theta$ and $\theta'_i=\Pi-\theta=\theta'$ we find
\begin{eqnarray}
E_W(\theta_1\theta_2\theta_3)&=&A(\cos\theta)^3+B(\sin\theta)^3+C\cos(3\theta)+ D(3(\cos\theta)^2\sin\theta),\nonumber \\
E_W(\theta'_1\theta'_2\theta'_3)&=&-A(\cos\theta)^3+B(\sin\theta)^3-C\cos(3\theta)+ D(3(\cos\theta)^2\sin\theta),\nonumber \\
E_W(\theta'_1\theta_2\theta_3)&=&-A(\cos\theta)^3+B(\sin\theta)^3-C\cos(\theta)-D((\cos\theta)^2\sin\theta)\nonumber
\\ &=& E_W(\theta_1\theta'_2\theta_3)= E_W(\theta_1\theta_2\theta'_3),\nonumber \\
E_W(\theta'_1\theta'_2\theta_3)&=&A(\cos\theta)^3+B(\sin\theta)^3+C\cos(\theta)-D((\cos\theta)^2\sin\theta)\nonumber
\\ &=& E_W(\theta_1\theta'_2\theta'_3)= E_W(\theta'_1\theta_2\theta'_3).
\end{eqnarray}
Thus for the inequalities, we get
\begin{eqnarray}\label{amin}
|M|&=&|-2A(\cos\theta)^3+2B(\sin\theta)^2-6D(\cos\theta)^2\sin\theta+ C(\cos3\theta-3\cos\theta)|,\nonumber\\
|M'|&=&|-2A(\cos\theta)^3-2B(\sin\theta)^2+6D(\cos\theta)^2\sin\theta+C(\cos3\theta-3\cos\theta)|,\nonumber\\
|S_v|&=&|M+M'|=|-4A(\cos\theta)^3 +2 C\cos3\theta -6C\cos\theta|.
\end{eqnarray}
In the non-relativistic limit ($\Omega_p\rightarrow0$) and by
using $\theta=35.264^{\circ}$, Eq.~(\ref{amin}) is consistent with
the measurements in the lab frame. In the high velocity limit
$(\beta\rightarrow1)$, we reach at
\begin{eqnarray}
A&\longrightarrow&-\frac{3}{2\Gamma^3}+\frac{7}{6\Gamma}\nonumber \\
B&\longrightarrow&\sqrt{1-\frac{1}{\Gamma^2}}(\frac{3}{\Gamma^2}-1)\nonumber \\
C&\longrightarrow&\frac{-3}{\Gamma^3}+\frac{7}{3\Gamma}\nonumber \\
D&\longrightarrow&\sqrt{1-\frac{1}{\Gamma^2}}(\frac{-3}{\Gamma^2}+\frac{2}{3}),
\end{eqnarray}
and finally we get
\begin{eqnarray}\label{awi}
|M|&\sim&|\frac{9.797}{\Gamma^3}-\frac{7.620}{\Gamma}+
1.14\sqrt{1-\frac{1}{\Gamma^2}}(\frac{9}{\Gamma^2}-2.19)|,\nonumber \\
|M'|&\sim&|\frac{9.797}{\Gamma^3}-\frac{7.620}{\Gamma}-
1.14\sqrt{1-\frac{1}{\Gamma^2}}(\frac{9}{\Gamma^2}-2.19)|,\nonumber \\
|S_v|&\sim&\vert\frac{19.594}{\Gamma^3}-\frac{15.236}{\Gamma}\vert,
\end{eqnarray}
for the inequalities. In addition, for the spin state in the
moving frame ($\vert W\rangle^\Lambda$), in the $\beta\rightarrow
1$ limit and the low energy regime ($\Gamma\rightarrow1$), we
reach
\begin{eqnarray}\label{w1}
\vert W \rangle^\Lambda\sim\vert W \rangle,
\end{eqnarray}
where for the high energy particles ($\Gamma\rightarrow\infty$) we get
\begin{eqnarray}\label{w2}
\vert W
\rangle^\Lambda\sim\frac{1}{2}(\sqrt{3}|\overline{GHZ}\rangle+\frac{1}{\sqrt{2}}
(\vert \overline{W}\rangle-\vert W\rangle)).
\end{eqnarray}

Eq.~(\ref{awi}) shows that, in the $\beta \rightarrow 1$ limit,
the inequalities in the moving frame are violated to their
violation value in the $S$ frame for the low energy particles
($\Gamma\rightarrow1$). This result is supported by the asymptotic
behavior of the spin state of system in this limit (\ref{w1}). In
addition, if we use Czachor's relativistic spin operator instead
of the Pauli spin operator then the inequalities will be satisfied
in this limit \cite{mm} leading to a full contradiction with
Eq.~(\ref{w1}). Therefore, it seems that the predictions of the
$|M|$ and $|M'|$ inequalities considering the Pauli spin operator
are in more agreement with the behavior of the entangled spin
state transformed by the LT compared with those in which Czachor's
spin operator are considered.

Let us note that the $|S_v|$ inequality for the high energy
particles ($\Gamma\rightarrow \infty$) is satisfied in the $\beta
\rightarrow 1$ limit (\ref{awi}). In addition, in the $\beta
\rightarrow 1$ limit, the violation amount of the $|M|$ and $|M'|$
inequalities for the high energy particles are more than those of the
low energy particles (\ref{awi}). Therefore, there is a
contradiction between the result of $|S_v|$ and those of $|M|$ and
$|M'|$.

In order to solve this contradiction, we evaluate
\begin{eqnarray}\label{wwb}
E(ABC)_{W\overline{W}}&=&\langle W|\sigma(\widehat{N_1})\otimes
\sigma(\widehat{N_2})\otimes \sigma(\widehat{N_3})
|\overline{W}\rangle
=-\frac{2}{3}\sin(\theta_1+\theta_2+\theta_3)\nonumber \\
&+&\frac{1}{3}\sin\theta_1 \sin\theta_2 \sin\theta_3 =\langle
\overline{W}|\sigma(\widehat{N_1})\otimes
\sigma(\widehat{N_2})\otimes \sigma(\widehat{N_3}) |W\rangle,
\end{eqnarray}
where $\widehat{N_i}$ are the same vectors used in order to evaluate
the correlation functions ($E_{W}(ABC)$) of the W state.
If we use $\vert \tau \rangle=\frac{1}{\sqrt{2}}(\vert
\overline{W}\rangle-\vert W\rangle)$ state instead of the W state in Eq.~(\ref{cf2}),
simple calculations lead to
\begin{eqnarray}\label{crns}
E(ABC)_{\tau}=E_{W}(ABC)-E(ABC)_{W\overline{W}}
\end{eqnarray}
and
\begin{eqnarray}\label{delta}
\Delta E(ABC)&=&E(ABC)_{\tau}-E_{W}(ABC)\nonumber \\
&=&-\frac{1}{3}\sin\theta_1
\sin\theta_2\sin\theta_3+\frac{2}{3}\sin(\theta_1+\theta_2+\theta_3).
\end{eqnarray}
Bearing $\theta_i=\pi-\theta'_i=35.264^{\circ}$ in mind, we get
$\Delta E(ABC)>0$ for the measurements ($A,A',B,...$). This
analysis shows that the particles in the $\vert \tau \rangle$
state are more entangled than the particles in the W state and
therefore, the violation amount of the inequalities for the $\vert
\tau \rangle$ state should be more than those of the W state. Once
again, we note that although the high energy limit is problematic
in the Quantum Mechanics framework, it is at least mathematically
useful to study this limit.

Comparing Eqs.~(\ref{w1}) and~(\ref{w2}), we conclude that the
violation amount of the inequalities should be increased as a
function of the energy of particles. Therefore, Eq.~(\ref{w2}) is
in agreement with the predictions of the $|M|$ and $|M'|$
inequalities for high energy particles while the $|S_v|$
inequality fails to predict this behavior (\ref{awi}). This result
indicates that the $|S_v|$ inequality is not a good witness for
studying the non-locality stored in the W state in the
relativistic regimes. From Eqs.~(\ref{w2}) and~(\ref{gbar}), it is
apparent that if the moving frame observer applies the
measurements used to detect the non-locality of the GHZ state, the
$|S_v|$ inequality will be violated to its maximum violation
value, this is due to the fact that the $\overline{\textmd{GHZ}}$
state is produced in this limit. In addition, this result shows
that the multi-particle Bell-like inequalities are more sensitive
to the directions of measurements in the relativistic regimes
compared with the Bell's inequality which is in agreement with the
relativistic \cite{mm} and non-relativistic studies \cite{cerc}.
Bearing the differences between the entanglement of the W state
and that of the GHZ state together with the results of previous
subsection in mind, it seems that the behavior of the W state
differs from that of the Bell states \cite{terashima,terashima1}
and the GHZ state \cite{you}, this shows that the behavior of
non-locality depends on both the number of the entangled particles
and their type of entanglement. If one uses Czachor's relativistic
spin operator instead of the Pauli spin operator then, in the
$\beta\rightarrow1$ limit, the inequalities will be satisfied
independent of the particles energy \cite{mm} which is again
indicating that the inequalities considering the Pauli spin
operator have better agreement with the behavior of the entangled
spin state affected by the LT compared with those in which
Czachor's spin operator are considered.
\section{Summary and Conclusion \label{Summary}}
The behavior of the GHZ and W states under the LT is investigated.
In addition, considering the Pauli spin operator, we studied the
behavior of three well-known classes of the multi-partite
Bell-like inequalities including the $|S_v|$, $|M|$ and $|M'|$
inequalities under the LT. We compared our results with those in
which Czachor's relativistic spin operator is used instead of the
Pauli operator. We used the behavior of the spin state under LT as
criterion in order to choose the appropriate inequalities and the
proper spin operator. In our setup, the moving and lab frames use
the same set of measurements violating the $|S_v|$ inequality to
its maximum violation amount ($|S_v|_m$) in the lab frame. By
these measurements, the $|M|$ and $|M'|$ inequalities are violated
to the same value ($\frac{|S_v|_m}{2}$) in the lab frame
simultaneously. Our results indicate that: ($\textmd{i}$) Since
the predictions of the $|S_v|$, $|M|$ and $|M'|$ inequalities,
when the Pauli operator is considered as the spin operator, are
compatible with the asymptotic behavior of the spin states in the
moving frame, the Pauli operator is more compatible with LT of the
quantum mechanical spin state compared with Czachor's spin
operator. This result is in full agreement with the two and
one-particle studies \cite{mgb}. ($\textmd{ii}$) The general
behavior of the GHZ and W states under the LT, in the
$\beta\rightarrow1$ limit, is independent of the type of
entanglement of the entangled particles if one considers the
$|S_v|$ inequality and the special set of measurements violating
$|S_v|$ to its maximum violation amount in the lab frame. It is
due to the fact that, in the $\beta\rightarrow1$ limit, the
violation amount of the $|S_v|$ inequality in the moving frame is
equal to that of the lab frame for the low energy particles
whereas it is not violated for the high energy particles.
Moreover, we found that the $|S_v|$ inequality can be considered
as a reasonable witness for studying the non-locality stored in
the GHZ state in the relativistic regime. Also, our study shows
that the inequalities with correlation functions stronger than
those of the $|M|$ and $|M'|$ inequalities, and weaker than those
of $|S_v|$ can be used in order to study the effects of LT on the
GHZ and W states supported by the non-relativistic study
\cite{AA}. ($\textmd{iii}$) The predictions of the $|M|$ and
$|M'|$ inequalities are always consistent with the asymptotic
behavior of spin state of the system if the lab and moving frames
use the special set of measurements violating $|S_v|$ to its
maximum violation amount in the lab frame. These behaviors show
that the $|M|$ and $|M'|$ inequalities can be considered as
reasonable inequalities for studying the effects of LT on the W
state. ($\textmd{iv}$) Since the behavior of the W state under the
LT differs from those of the GHZ and Bell states, it seems that
the behavior of non-locality depends on the number of the
entangled particles and their type of the entanglement. This
result is supported by the behavior of the inequalities under the
LT. ($\textmd{v}$) The violation amount of the $|S_v|$, $|M|$ and
$|M'|$ inequalities for the low energy particles measured in the
moving frame, in the $\beta\rightarrow 1$ limit, is equal to that
of the lab frame. The same result is reported in the case of the
two-particle systems \cite{terashima,terashima1}. Therefore, we
note that this result is independent of the number of the
entangled particles and their type of entanglement. Finally, we
should note that a Stern-Gerlach type experiment is needed for
investigating our results.

By comparing our results with the results reported in
\cite{moradi1,mm,moradi2,you}, we can conclude that the
sensitivity of the three-particle Bell-like inequalities to the
set of measurements in the relativistic regimes is more than that
of the Bell's inequality \cite{terashima,terashima1,ahn,lee,kim}
which is in line with the non-relativistic \cite{cerc} and the
relativistic studies \cite{mm}.
\section*{Acknowledgements}
We are grateful to the anonymous reviewers for their worthy hints
and constructive comments which help us increase our understanding
of the subject. This work has been supported financially by
Research Institute for Astronomy and Astrophysics of Maragha
(RIAAM) under research No.$1/3720-76$.


\begin{thebibliography}{99}
\bibitem{EPR} Einstein, E., Podolsky, B., Rosen, N., Phys. Rev. \textbf{47}, 777 (1935).
\bibitem{BA} Bohm, D., Aharanov, Y., Phys. Rev. \textbf{108}, 1070 (1957).
\bibitem{Bell} Bell, J.S., Physics (N.Y). {\bf1}, 195 (1964).
\bibitem{CHSH} Clauser, J.F., Horne, M.A., Shimony, A., Holt, R.A., Phys. Rev. Lett. {\bf23}, 880 (1969).
\bibitem{WIG} Wigner, E.P., Am. J. Phys. {\bf38}, 1005 (1970).
\bibitem{CH} Clauser, J.F., Horne, M.A., Phys. Rev. D {\bf10}, 526 (1974).
\bibitem{rev} Brunner, N., Cavalcanti, D., Pironio, S., Scarani, V., Wehner, S., Rev. Mod. Phys. {\bf86}, 419 (2014).
\bibitem{aspect1} Aspect, A., Grangier, P., Roger, G., Phys. Rev. Lett. {\bf47}, 460 (1981).
\bibitem{aspect2} Aspect, A., Grangier, P., Roger, G., Phys. Rev. Lett. {\bf49}, 91 (1982).
\bibitem{aspect3} Aspect, A., Grangier, P., Roger, G., Phys. Rev. Lett. {\bf49}, 1804 (1982).
\bibitem{vedral1} Dunningham, J.A., Vedral, V., Phys. Rev. Lett. {\bf99}, 180404 (2007).
\bibitem{vedral2} Cooper, J.J., Dunningham, J.A., New J. Phys. {\bf10}, 113024 (2008).
\bibitem{gisin91} Gisin, N., Phys. Lett. A {\bf154}, 201 (1991).
\bibitem{gisinperes} Gisin, N., Peres, A., Phys. Lett. A {\bf162}, 15 (1992).
\bibitem{pop92} Popescu, S., Rohrlich, D., Phys. Lett. A {\bf166}, 293 (1992).
\bibitem{werner89} Werner, R.F., Phys. Rev. A {\bf40}, 4277 (1989).
\bibitem{gisin96} Gisin, N., Phys. Lett. A {\bf210}, 151 (1996).
\bibitem{eber} Eberhard, P., Phys. Rev. A {\bf47}, 747(R) (1993).
\bibitem{eber1} Acin, A., Durt, T., Gisin, N., Latorre, J.I., Phys. Rev. A {\bf65}, 052325 (2002).
\bibitem{eber2} Acin, A., Gill, R., Gisin, N., Phys. Rev. Lett. {\bf95}, 210402 (2005).
\bibitem{eber3} Zohren, S., Gill, R.D., Phys. Rev. Lett. {\bf100}, 120406 (2008).
\bibitem{jun} Junge, M., Palazuelos, C., Commun. Math. Phys. {\bf306}, 695 (2011).
\bibitem{eber4} Vidick, T., Wehner, S., Phys. Rev. A {\bf83}, 052310 (2011).
\bibitem{ben} Bennett, C.H., Divincenzo, D.P., Fuchs, C.A., Mor, T., Rains, E., Shor,
P.W., Wootters, W.K., Phys. Rev. A {\bf59}, 2 (1999).
\bibitem{Bert} Bertlmann, R.A., J. Phys. A: Math. Theor. {\bf47}, 424007 (2014).
\bibitem{aud} Audretch, J.,: Entangled systems. Willy-VCH verlag, (2007).
\bibitem{Nilsen} Nielsen, N.A., Chuang, I.L.,: Quantum Computation and Quantum Information.
Cambridge University Press, Cambridge, England, (2000).
\bibitem{wald} Wald, R.M.,: Quantum field theory in Curved Spacetime and Black Hole
Thermodynamics. University of Chicago Press, Chicago, (1994).
\bibitem{peres1} Peres, A., Scudo, P.F., Terno, D.R., Phys. Rev. Lett. {\bf88}, 230402 (2002).
\bibitem{vedral3} Dunningham, J., Palge, V., Vedral, V., Phys. Rev. A {\bf80}, 044302 (2009).
\bibitem{peres2} Cezachor, M., Phys. Rev. Lett. {\bf94}, 078901 (2005).
\bibitem{peres3} Peres, A., Scudo, P.F., Terno, D.R., Phys. Rev. Lett. {\bf94}, 078902 (2005)
\bibitem{peres4} Czachor, M., Wilczewski, M., Phys. Rev. A {\bf68}, 010302(R) (2003).
\bibitem{peres5} Czachor, M., arXiv:quant-ph/0205187v1.
\bibitem{wein} Weinberg, S.,: The Quantum Theory of Fields. Volume I: Foundations, Cambridge University
Press, Cambridge (1995).
\bibitem{fun} Fuentes, I., Mann, R. B., Martin-Martinez, E., Moradi, S., Phys. Rev. D {\bf82}, 045030 (2010).
\bibitem{smith} Smith, A., Mann, R.B., Phys. Rev. A {\bf86}, 012306 (2012).
\bibitem{friis} Friis, N., K\"{o}hler, P., Martin-Martinez, E., Bertlmann, R.A., Phys. Rev. A {\bf 84}, 062111 (2011).
\bibitem{moradi3} Moradi, S., JETP Letters. {\bf89}, 1 (2009).
\bibitem{moradi4} Moradi, S., Pierini, R., Mancini, S., Phys. Rev. D {\bf89}, 024022 (2014).
\bibitem{hal} Halpern, F.R.,: Special Relativity and Quantum Mechanics. Prentice-Hall, Englewood Cliffs, NJ, (1968).
\bibitem{wigner} Wigner, E., Ann. Math. {\bf40}, 149 (1939).
\bibitem{gingrich} Gingrich, R.M., Adami, C., Phys. Rev. Lett. \textbf{89}, 270402 (2002).
\bibitem{li} Li, H., Du, J., Phys. Rev. A {\bf68}, 022108 (2003).
\bibitem{jordan} Jordan, T.F., Shaji, A., Sudarshan, E.C.G., Phys. Rev. A {\bf75}, 022101 (2007).
\bibitem{alsing} Alsing, P.M., Milburn, G.J., Quant. Inf. Comput. {\bf2}, 487 (2002).
\bibitem{terashima} Terashima, H., Ueda, M., Quant. Inf. Comput. {\bf3}, 224 (2003).
\bibitem{terashima1} Terashima, H., Ueda, M., Int. J. Quant. Inf. {\bf1}, 93 (2003).
\bibitem{ahn} Ahn, D., Lee, H-J., Moon, Y.H., Hwang, S.W., Phys. Rev. A {\bf67}, 012103 (2003).
\bibitem{lee} Lee, D., Chang-Young, E., New J. Phys. {\bf6}, 67 (2004).
\bibitem{kim} Kim, W.T., Son, E.J., Phys. Rev. A {\bf71}, 014102 (2005).
\bibitem{fu} Fuentes-Schuller, I., Mann, R.B., Phys. Rev. Lett. {\bf95}, 120404 (2005).
\bibitem{al} Alsing, P.M., Fuentes-Schuller, I., Mann, R.B., Tessier, T.E., Phys. Rev. A {\bf74}, 032326 (2006).
\bibitem{ma} Mann,  R.B., Villalba, V.M., Phys. Rev. A {\bf80}, 022305 (2009).
\bibitem{le} Leon, J., Martin-Martinez, E., Phys. Rev. A {\bf80}, 012314 (2009).
\bibitem{tera} Terashima, H., Ueda, M., Phys. Rev. A {\bf69}, 032113 (2004).
\bibitem{shi} Shi, Y., Phys. Rev. D {\bf70}, 105001 (2004).
\bibitem{ball} Ball, J.L., Schuller, I.F., Schuller, F.P., Phys. Lett. A {\bf359}, 550 (2006).
\bibitem{ver} Ver Steeg, G., Menicucci, N.C., Phys. Rev. D {\bf79}, 044027 (2009).
\bibitem{czachor} Czachor, M., Phys. Rev. A {\bf55}, 72 (1997).
\bibitem{spin1} Bauke, H., Ahrens, S., Keitel, C.H., Grobe, R. Phys. Rev. A {\bf89}, 052101 (2014).
\bibitem{spin2} Bauke, H., Ahrens, S., Keitel, C.H., Grobe, New. J. Phys. {\bf16}, 043012 (2014).
\bibitem{spin4} Terno, D.R., Phys. Rev. A {\bf67}, 014102 (2003).
\bibitem{moradi1} Moradi, S., Phys. Rev. A {\bf77}, 024101 (2008).
\bibitem{mm} Moradpour, H., Montakhab, A., under review in Phys. Rev. A.
\bibitem{moradi2} Moradi, S., Aghaee, M., Int. J. Theor. Phys. {\bf49}, 615 (2010).
\bibitem{genu} Ghose, S., Sinclair, N., Debnath, S., Rungta, P., Stock, R., Phys. Rev. Lett. {\bf102}, 250404 (2009).
\bibitem{genu1} Ghose, S., Debnath, S., Sinclair, N., Kabra, A., Stock, R., J. Phys. A {\bf43}, 445301 (2010).
\bibitem{dur} Dur, W., Vidal, G., Cirac, J.I., Phys. Rev. A {\bf62}, 062314 (2000).
\bibitem{Mitchel} Mitchell, P., Popescu, S., Roberts, D., Phys. Rev. A {\bf70}, 060101(R) (2004).
\bibitem{cerc} Cereceda, J. L., Phys. Rev. A {\bf66}, 024102 (2002).
\bibitem{svet} Svetlichny, G., Phys. Rev. D {\bf35}, 3066 (1987).
\bibitem{svet1} Seevinck, M., Svetlichny, G., Phys. Rev. Lett. {\bf89}, 060401 (2002).
\bibitem{cofman} Coffman, V., Kundu, J., Wootters, W.K., Phys. Rev. A {\bf61}, 052306 (2000).
\bibitem{mermin} Mermin, N.D., Phys. Rev. Lett. {\bf65}, 1838 (1990).
\bibitem{cal} Collins, D., Gisin, N., Popescu, S., Roberts D., Scarani, V., Phys. Rev. Lett. {\bf88}, 170405 (2002).
\bibitem{gisin} Gisin, N., Bechmann-Pasquinucci, H., Phys. Lett. A {\bf246}, 1 (1998).
\bibitem{roy} Roy, S.M., Phys. Rev. Lett. {\bf94}, 010402 (2005).
\bibitem{AA} Ajoy, A., Rungta, P., Phys. Rev. A {\bf81}, 052334 (2010).
\bibitem{bp} Bancal, J.-D., Barrett, J., Gisin, N., Pironio, S., Phys. Rev. A {\bf88}, 014102 (2013).
\bibitem{gw} Gallego, R., W$\ddot{u}$rflinger, L.E., Ac\'{i}n, A., Navascu$\acute{e}$s, M., Phys. Rev. Lett.
{\bf109}, 070401 (2012).
\bibitem{ac} Almeida, M., Cavalcanti, D., Scarani, V., Ac\'{i}n, A., Phys. Rev. A {\bf81}, 052111 (2010).
\bibitem{you} You, H., Wang, A.M., Yang, X., Niu, W., Ma, X., Xu, F., Phys. Lett. A {\bf333}, 389 (2004).
\bibitem{hwang} Hwang, M., Park, D., Jung, E., Phys. Rev. A {\bf83}, 012111 (2011).
\bibitem{wang} Wang, J., Jing, J., Phys. Rev. A {\bf83}, 022314 (2011).
\bibitem{mgb} Moradpour, H., Bahadoran, M., arXiv:1506.07106.
\end{thebibliography}
\end{document}